\documentclass[twocolumn,amsmath,amssymb]{revtex4}
\usepackage[english]{babel}
\usepackage[dvips]{graphicx}
\usepackage{enumerate}
\begin{document}
\title{Dark Matter annihilations in Pop III stars}
\author{Marco Taoso$^{1,2}$}
\author{Gianfranco Bertone$^{2}$}
\author{Georges Meynet$^{3}$}
\author{Silvia Ekstr$\ddot{\mbox{o}}$m$^{3}$}
\affiliation{$^{1}$ INFN, Sezione di Padova, via Marzolo
8,Padova,35131,Italy} \affiliation{$^{2}$ Institut d'Astrophysique
de Paris, UMR 7095-CNRS, Universit\'e Pierre et Marie Curie, 98 bis
Boulevard Arago 75014, Paris, France} \affiliation{$^{3}$ Geneva
Observatory, University of Geneva, Maillettes 51, 1290 Sauverny,
Switzerland}

\begin{abstract}
We study the impact of the capture and annihilation of Weakly
Interacting Massive Particles (WIMPs) on the evolution of Pop
III stars. With a suitable
modification of the Geneva stellar
evolution code, we study the evolution of 20 and 200 M$_\odot$
stars in Dark Matter haloes with densities between 10$^{8}$ and
$10^{11}$ GeV/cm$^3$ during the core H-burning phase, and, for selected
cases, until the end of the core He-burning phase.
We find that for WIMP densities higher than 5.3 $10^{10}(\sigma^{SD}_p/10^{-38} \mbox{ cm}^2)^{-1}$ GeV cm$^{-3}$
the core H-burning lifetime of $20 M_{\odot}$ and $200 M_{\odot}$ stars
exceeds the age of the Universe, and stars are sustained only by
WIMP annihilations. We determine the observational properties of
these `frozen` objects and show that they can be searched for in the local
Universe thanks to their anomalous mass-radius relation, which should
allow unambiguous discrimination from normal stars.
\end{abstract}

\maketitle

In the Standard Cosmological Model, the matter density of the
Universe is dominated by an unknown component, approximately 5 times
more abundant than baryons, dubbed Dark Matter (DM). Among the many
DM candidates proposed in the literature, Weakly Interacting Massive
Particles (WIMPs), i.e. particles with mass $\cal{O}$$(100)$ GeV and
weak interactions, appear particularly promising, also in view of
their possible connection with well motivated extensions of the
Standard Model of particle physics (see Ref.~\cite{reviews} for
recent reviews on particle DM, including a discussion of ongoing
direct, indirect and accelerator searches).
Despite their weak interactions, WIMPs can lead to macroscopic
effects in astrophysical objects, provided that they have a sizeable
scattering cross section off baryons. In this case, in fact, DM
particles traveling through stars can be captured, and sink at the
center of the stars. Direct searches and astrophysical arguments,
however, severely constrain the strength of DM-baryons interactions
(see e.g. Ref.~\cite{mack} and references therein). Since the
capture rate is proportional to the product of the scattering cross
section times the local DM density, large effects are thus expected
in regions where the DM density is extremely high (this was already
noticed in the context of the so called 'cosmions' \cite{cosmions}).
Recent progress in our understanding of the formation and structure
of DM halos has prompted a renewed interest in the consequences of
DM capture in stars, in particular in the case of White
Dwarfs~\cite{Moska}, compact objects~\cite{Bertone:2007ae} and main
sequence stars~\cite{Fairbairn} at the Galactic center, where the DM
density could be extremely high~\cite{Bertone:2005hw}.

Alternatively, one may focus on the first stars, which are thought to
form from gas collapsing at the center of $10^6-10^8 M_{\odot}$
DM halos at redshift $z\lesssim 10-30$.
In fact, Spolyar, Freese and Gondolo \cite{Freese} have shown that
the energy released by WIMPs annihilations in these mini-halos,
during the formation of a proto-star (thus even before DM capture
becomes efficient), may exceed any cooling mechanism, thus
inhibiting or delaying stellar evolution (see also Ref.~\cite{Freeseb}).
The formation of proto-stars with masses
between $6 M_{\odot}$ and $600 M_{\odot}$ in DM halos of $10^6
M_{\odot}$ at z=20, can actually be delayed by $\sim 10^3-10^4 $ yrs~\cite{Iocco}. Once the
star forms, the scattering of WIMPs off the stellar nuclei becomes
more efficient and a large number of WIMPs can be trapped inside the
gravitational potential well of the star. The WIMPs luminosity can
overwhelm that from nuclear reactions and therefore strongly modify
the star evolution~\cite{Iocco2,Freese2}, and the core H-burning
phase of Pop III stars, in DM
halos of density of $10^{11} \mbox{ GeV cm}^{-3}$, is substantially prolonged,
especially for small mass stars ($M_{*}<40 M_{\odot}$)~\cite{Iocco}.

In this letter, we perform a detailed study of the impact of DM
capture and annihilation on the evolution of Pop. III stars with a suitable
modification of the Geneva stellar
evolution code~\cite{Ekstrom Maynet}. With respect to previous analyses,
this already allows us to properly take into account the stellar
structure in the calculation of the capture rate, that we compute, following Ref.~\cite{Gould}, as
\begin{figure}[t]
  \resizebox{\hsize}{!}{\includegraphics[width=4.5 cm]{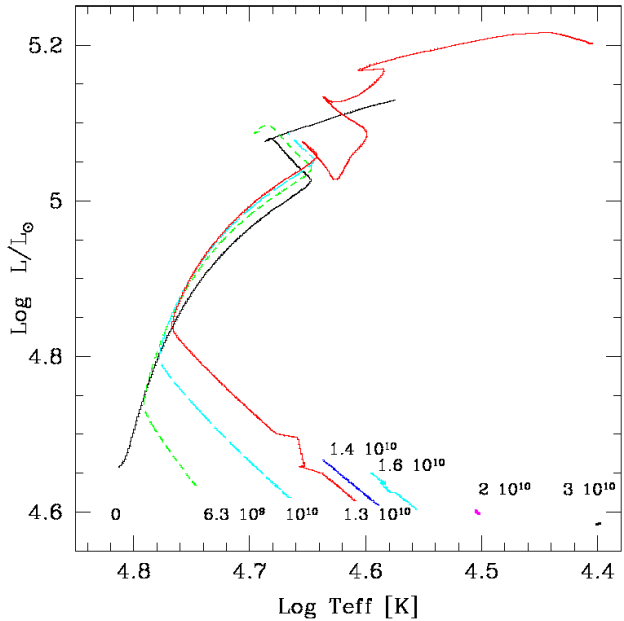}}
  \caption{Evolutionary tracks of a Pop III 20 M$_\odot$ star for different
WIMP densities (labels in units of GeV cm$^{-3}$).
We have adopted a WIMP model with $m_{\chi}=100$ GeV and $\sigma^{SD}_p =10^{-38}$
cm$^2$.}
  \label{HR}
\end{figure}
\begin{equation}
C= 4 \pi \int_{0}^{R_*} dr r^2 \frac{dC(r)}{dV} \label{eqn:C}
\end{equation}
with
\begin{eqnarray}
\frac{dC(r)}{dV} &=&  \left(\frac{6}{\pi}\right)^{1/2}
\sigma_{\chi,N} \frac{\rho_{i}(r)}{M_i}\frac{\rho_{\chi}}{m_{\chi}}
\frac{v^{2}(r)}{\bar{v}^2} \frac{\bar{v}}{2 \eta A^2} \\
\nonumber &\times & \left\{ \left( A_+ A_- -\frac{1}{2}\right)
[\chi(-\eta,\eta)-\chi(A_-,A_+) ] \right.\\ \nonumber &+& \left.
\frac{1}{2} A_+ e^{-A_-^2} -\frac{1}{2} A_- e^{-A_+^2} -\frac{1}{2}
\eta e^{-\eta^2} \right\} \label{eqn:dCdV}
\end{eqnarray}
$$ A^2=\frac{3 v^2(r)\mu}{2 \bar{v}^2 \mu_-^2} \mbox{,  }\hspace{0.5cm}
A_{\pm}=A \pm \eta \mbox{,}
\hspace{0.5cm}\eta^2=\frac{3v_{*}^2}{2\bar{v}^2}$$
$$\chi(a,b)=\frac{\sqrt{\pi}}{2}[\mbox{Erf}(b)-\mbox{Erf}(a)]=\int_a^bdy e^{-y^2}$$
$$\mu_-=(\mu_i-1)/2 \mbox{,} \hspace{0.5cm} \mu_{i}=m_{\chi}/M_i$$
where $\rho_i(r)$ is the mass density
profile of a given chemical element in the interior of the star and
$M_i$ refers to its atomic mass, while $\rho_{\chi}$, $m_{\chi}$ and
$\bar{v}$ are respectively the WIMP mass and the WIMP density and
velocity dispersion at the star position. The velocity of the star
with respect to an observer, labeled as $v_*$, is assumed to be
equal to $\bar{v}$, giving therefore $\eta=\sqrt{3/2}$. The radial
escape velocity profile depends on $M(r)$, i.e. the mass enclosed within a
radius $r$, $v^2(r)=2 \int_{r}^{\infty} G M(r^{'})/r^{'2} dr'$.
\begin{figure}[t]
  \resizebox{\hsize}{!}{\includegraphics[width=4.5 cm]{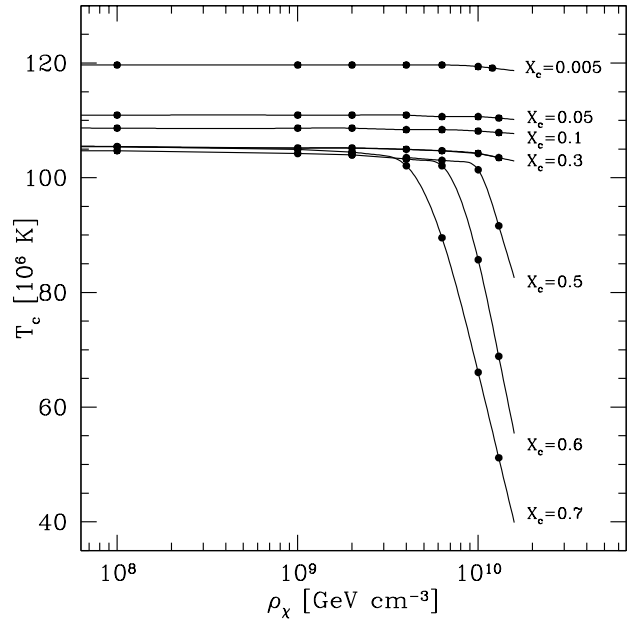}}
  \caption{Temperature of the core as a function of the DM density for
the 20 M$_\odot$ model, at different stages of the core H-burning phase.
$X_c$ denotes the mass fraction of hydrogen at the centre (
$X_c=0.76$ at the beginning of the core H-burning phase). WIMP parameters as in Fig. 1.} \label{tcxc}
\end{figure}
The WIMP scattering cross section off nuclei, $\sigma_{\chi,N}$ is
constrained by direct detection experiments and for a WIMP mass of
100 GeV the current upper limits are $\sigma_{SI}=10^{-43} \mbox{
cm}^2$~\cite{SI} and $\sigma_{SD}=10^{-38} \mbox{ cm}^2 $~\cite{SD} respectively for
spin-independent and spin-dependent WIMP interactions off a proton.
We will adopt these reference values throughout the paper, but the
capture rate can be easily rescaled for other scattering cross
sections by using Eq. \ref{eqn:dCdV}. The spin-independent
interactions with nucleons inside nuclei add up coherently giving
an enhancement factor $A^4$ with respect to the interaction with a
single nucleon: $\sigma^{SI}_{\chi,N}=A^4\sigma_{\chi,p}$, where $A$
is the mass number. There is no such enhancement for the
spin-dependent interactions.
We consider the contribution to the capture rate from WIMP-hydrogen
spin dependent interactions and WIMP-helium $^4$He spin-independent
interactions, neglecting the presence of other elements because of
their very low abundance. The contribution of Helium is found to be
negligible with respect to that from hydrogen.

Once captured, WIMPs get redistributed in the interior of the star
reaching, in a characteristic time $\tau_{th},$ a thermal
distribution \cite{GriestSeckel}:
\begin{equation}
n_{\chi}(r)=n_0 e^{\frac{-r^2}{r_w^2}} \mbox{ with }
r_{\chi}=\sqrt{\frac{3 k T_c}{2\pi G \rho_c m_{\chi}}}
\label{eq:Distribuzione}
\end{equation}

\noindent with $T_c$ and $\rho_c$ referring to the core temperature
and density. The distribution results quite concentrated toward the
center of the star: e.g. for a $20 M_{\odot}$ star immersed in a
WIMP density of $\rho_{\chi}=10^{9} \mbox{ GeV cm}^{-3}$ at the
beginning of the core H-burning phase we obtain $r_{\chi}= 2 \times
10^{9} \mbox{cm},$ a value much lower than the radius of the star,
$R_*= 10^{11} \mbox{cm}.$ This consideration underlines the
importance of an accurate spatial resolution in the core to properly
treat the luminosity produce from WIMPs annihilations. We have also
checked that regardless the extremely high concentrations of WIMPs
obtained at the center of the stars, the gravity due to WIMPs is
completely negligible.
\begin{figure}[t]
  \resizebox{\hsize}{!}{\includegraphics[width=4.5 cm]{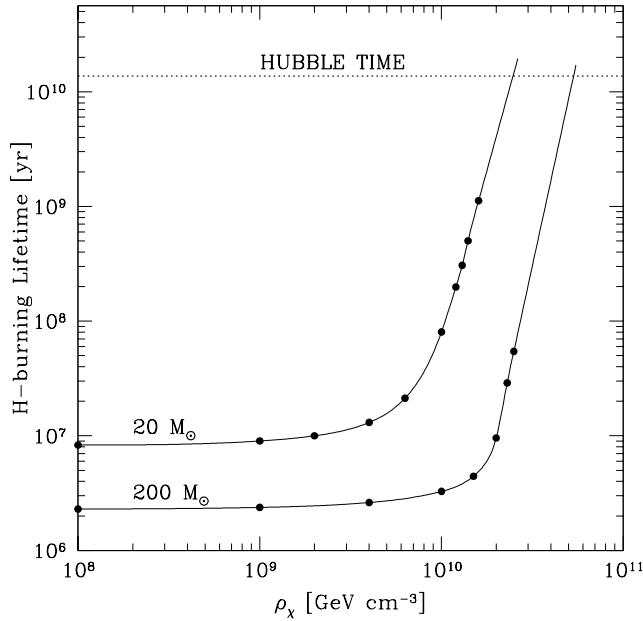}}
  \caption{Variation of the core H-burning lifetime as a function of the WIMP densities for the
  Pop III 20 and 200 M$_\odot$ models. WIMP parameters as in Fig. 1.}
  \label{den}
\end{figure}
\begin{figure}[t]
  \resizebox{\hsize}{!}{\includegraphics[width=4.5 cm]{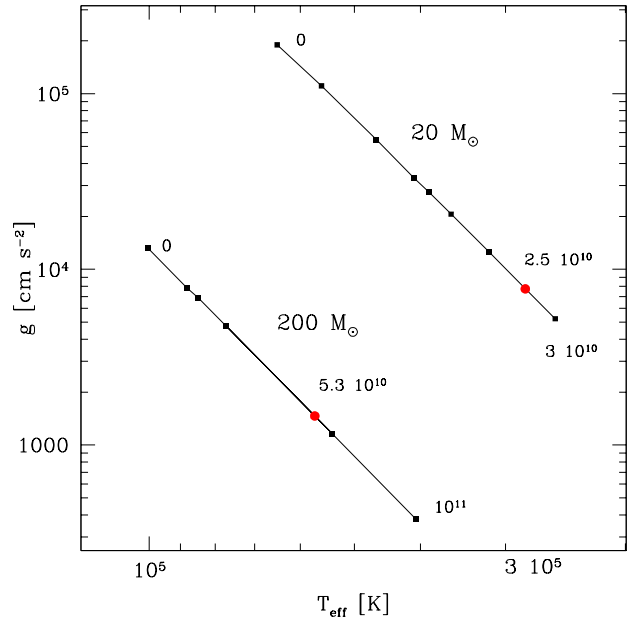}}
  \caption{ZAMS positions of 20 and 200 M$_\odot$ Pop. III stars in the g vs. $T_{eff}$ plane for different DM densities (labels in units of GeV cm$^{-3}$). Big red
circles correspond to the critical WIMP density (see text). WIMP parameters as in Fig. 1.}
  \label{gt}
\end{figure}
The number of scattering events needed for DM particles to thermalize with
the nuclei in the star is of
order $\approx m_{\chi}/M_H$, thus an upper limit on the thermalization
time can be obtained as $\tau_{th}=(m_{\chi}/M_H)/(\sigma_{SD}\bar{n}_H \bar{v})$ where $\bar{n_H}$ is the average
density on the star.

The WIMPs luminosity is simply $L_{\chi}(r)= 4 \pi (\sigma v) m_{\chi} c^2 n_{\chi}^2(r)$.
For the annihilation cross section
times relative velocity $(\sigma v)$, we assume the value $3 \times10^{-26}
\mbox{ cm}^2$, as appropriate for a thermal WIMP, but note that
the total WIMP luminosity at equilibrium does not depend on this
quantity. After a time
$$\tau_{\chi}= \left(\frac{C (\sigma v)}{\pi^{3/2} r_{\chi}^3} \right)^{-1/2}$$
an equilibrium between capture and annihilation is established, and
this incidentally allows to determine the normalization constant
$n_0$ above.

We have checked that the two transients $\tau_{\chi}$ and
$\tau_{th}$ remain much smaller, during the evolution of the star, than the
Kelvin-Helmotz timescale, $\tau_{KH}$ and the timescale needed for
the nuclear reactions to burn an hydrogen fraction $\Delta
X_c=0.002$ of the convective core, $\tau_{nucl}$:
$$\tau_{KH}=\frac{G M_*^2}{R_* L_*}\hspace{0.5cm} \tau_{nucl}=\frac{q_c \Delta X_c M_* 0.007 c^2}{L_*}$$
where the * labels quantities relative to the the star and $q_c$ is the
core convective mass fraction. This argument justifies the assumption
of equilibrium between capture and annihilation and the use of the
radial distribution in Eq. \ref{eq:Distribuzione}.
We assume here an average WIMP velocity
$\bar{v}= 10 \mbox{ Km s}^{-1},$ the virial velocity in an halo of
$10^{5}-10^6 M_{\odot}$ at z=20. As for the
DM density, semi-analytic computations of the adiabatic contraction
of DM halos \cite{Freese2, Freese3}, in agreement
with the results extrapolated from simulations of first star
formation \cite{AbelBryan}, suggest DM densities of order
$10^{12} \mbox{ GeV cm}^{-3}$ or even higher.


We have implemented the effects of WIMPs annihilation in the Geneva
stellar evolution code (see Ref.~\cite{Ekstrom Maynet} for details),
and followed the evolution of a $20 M_{\odot}$ and $200 M_{\odot}$
stars for different DM densities. We show in Fig.\ref{HR} the
evolutionary tracks for the $20 M_{\odot}$ model, and show for
comparison (black line) the case of a standard Pop III star without
WIMPs. For DM densities smaller than $10^9 \mbox{ GeV cm}^{-3}$ the
evolutionary tracks closely follow that of a normal star and they
are not shown for simplicity. The position of the star at the
beginning of the core H-burning phase (zero-age main sequence, or
ZAMS) is obtained when, after a short transient, the luminosity
produced at the center of the star equals the total luminosity and
the star settles down in a stationary regime. For increasing DM
densities the WIMPs luminosity produced at the center overwhelms the
luminosity from nuclear reactions and makes the star inflate,
producing therefore a substantial decrease of the effective
temperature and a moderate decrease of the star luminosity at the
ZAMS position, with respect to the standard scenario. For
$\rho_{\chi}=10^{10} \mbox{ GeV cm}^{-3}$, the energy produced by
WIMPs present in the star at a given time, estimated as $E_{\chi}
\simeq L_{\chi} \tau_{KH},$ is, at the ZAMS, $\sim 0.8$ times the
gravitational potential energy of the star, and the star therefore
starts to contract. In this phase, the core temperature, and
consequently also the nuclear reactions, increase. When the latter
become comparable with the WIMPs luminosity, the standard situation
is recovered and the evolutionary track joins the classical tracks
of a star without WIMPs. An important difference from standard
evolution is that in the first phase, the nuclear reactions are
slowed down and therefore the core H-burning lifetime is prolonged.
For Dark Matter densities $\rho_{\chi}\leq1.6$ $10^{10} \mbox{ GeV
cm}^{-3},$ the picture is qualitatively the same, and for these
models we only show in Fig.~\ref{HR} the first phases of the
evolution. In Fig.~\ref{tcxc}, we show the core temperature as a
function of the DM density, at different stages of the core
H-burning phase. At high DM densities hydrogen burns at much lower
core temperatures than in the usual scenario, till a certain mass
fraction is reached, e.g. $X_c=0.3$ for $\rho_{\chi}=10^{10} \mbox{
GeV cm}^{-3},$ and the standard evolutionary track is joined. For
increasing DM densities the nuclear reaction rate is more and more
delayed till the contraction of the star is inhibited, due to the
high DM energy accumulated, and the evolution is frozen. In
Fig.\ref{HR} for $\rho_{\chi}=2\cdot10^{10} \mbox{ GeV cm}^{-3}$ and
$\rho_{\chi}=3\cdot10^{10} \mbox{ GeV cm}^{-3}$ the stars seems to
remain indefinitely at the ZAMS position. In Fig.~\ref{den} we show
the core H-burning lifetime as a function of the DM density. In the
case of a $20 M_{\odot}$ model, for $\rho\leq10^{10} \mbox{ Gev
cm}^{-3}$ the core H-burning phase is prolonged by less then 10 \%
but the delay increases rapidly for higher DM densities.
Extrapolating the curve we determine a critical density, $\rho_c
=2.5\cdot10^{10} \mbox{ Gev cm}^{-3}$, beyond which the core
H-burning lifetime is longer then the age of the Universe. All the
calculations have been repeated for the $200 M_{\odot}$ model and we
find that both the $20 M_{\odot}$ and $200 M_{\odot}$ stars
evolutions are stopped for DM densities higher than $5.3\cdot10^{10}
(\frac{\sigma^{SD}_p}{10^{-38} \mbox{ cm}^2})^{-1} \mbox{ GeV
cm}^{-3}.$ We have also verified that the results weakly depend on
the WIMP mass, e.g. the core H-lifetime is modified by a factor
0.2\% and 5\% respectively for $m_{\chi}=10 \mbox{ GeV}$ and
$m_{\chi}=100 \mbox{ GeV},$ if $\rho_{\chi}=10^{10}\mbox{ GeV
cm}^{-3}$.

It is remarkable that under these circumstances, frozen Pop III stars
can survive until the present epoch, and can be searched for as an
anomalous stellar population. In Fig.\ref{gt} we show
the effective temperature and gravity acceleration at the surface of
these frozen Pop III stars, kept in the H-burning phase, for
different DM densities. Frozen stars would thus appear much
bigger and with much lower surface temperatures with respect to
normal stars with the same mass and metallicity.
Our results are qualitatively consistent with the preliminary
estimates in \cite{Iocco2, Freese2} and the analysis in
\cite{Iocco}. However, for a given DM density, we obtain a somewhat
longer core H-burning lifetime with respect to \cite{Iocco}, possibly due
to their use of an approximated expression for the capture rate.
We have also followed, for selected models, the evolution during the
core He-burning phase. During this evolutionary stage, the Dark
Matter luminosity is lower than the nuclear reaction luminosity,
therefore the impact of DM annihilations is found to be
rather weak. For the $20 M_{\odot}$ model and for $\rho_{\chi} =
1.6\cdot10^{10} \mbox{ GeV cm}^{-3}$ the He-lifetime is prolonged by a
factor 1.2, rather than a factor 37 found for the H-burning phase
for the same DM density.

In conclusion, we have adapted a stellar evolution code to the
study the evolution of Pop. III stars in presence of WIMPs. We have shown
that above a critical DM density, the
annihilation of WIMPs {\it captured} by Pop. III stars can dramatically
alter the evolution of these objects,
and prolong their lifetime beyond the age of the Universe. We have
determined the properties of these 'frozen' stars, and determined
the observational properties that may allow to discriminate these
objects from ordinary stars.

M.T. thanks the International Doctorate on Astroparticle Physics
(IDAPP) for partial support and the Geneva Observatory for the warm
hospitality. We thank F. Iocco for useful discussions.
During the completion of this work we became aware of a related
work done independently by Yoon, Iocco and Akiyama~\cite{Yoon}.
Their results, obtained with an independent stellar evolution code,
appear to be in good agreement with our own.


\begin{thebibliography}{99}
\bibitem{reviews} G. Bertone, D. Hooper and J. Silk, Phys. Rep. 405
(2005) 279; L.~Bergstrom,  Rept.\ Prog.\ Phys.\  {\bf 63}, 793 (2000)
\bibitem{mack}
  G.~D.~Mack, J.~F.~Beacom and G.~Bertone,
  Phys.\ Rev.\  D {\bf 76} (2007) 043523
  [arXiv:0705.4298 [astro-ph]].
\bibitem{cosmions} P. Salati and J. Silk, ApJ 296, 679 (1985); A. Renzini, Astr. Ap. 171, 121 (1987); A. Bouquet and P. Salati, ApJ 346, 284 (1989);D. Dearborn, G. Raffelt, P. Salati, J. Silk and A. Bouquet, ApJ 354, 568 (1990);P. Salati, G. Raffelt and D. Dearborn, ApJ 357, 566 (1990)
\bibitem{Moska} I.V. Moskalenko and L.L. Wai, Astrophys. J. 659:L29-L32, 2007 [arXiv:astro-ph/0702654].
\bibitem{Bertone:2007ae}
  G.~Bertone and M.~Fairbairn,
  Phys.\ Rev.\  D {\bf 77} (2008) 043515
  [arXiv:0709.1485 [astro-ph]].
\bibitem{Fairbairn} M. Fairbairn, P. Scott and J. Edsjo, Phys. Rev.
D 77 (2008) 047301
\bibitem{Bertone:2005hw}
  G.~Bertone and D.~Merritt,
  Phys.\ Rev.\  D {\bf 72} (2005) 103502
  [arXiv:astro-ph/0501555].
\bibitem{Freese} D. Spolyar, K. Freese and P. Gondolo, Phys. Rev.
Lett. 100 (2008) 051101.
\bibitem{Freeseb} K. Freese, P. Bodenheimer, D. Spolyar and P.
Gondolo, 2008 [arXiv:0806.0617].
\bibitem{Iocco} F. Iocco, A. Bressan, E. Ripamonti, R. Schneider, A.
Ferrara and P. Marigo, 2008 [arXiv:0805.4016].
\bibitem{Iocco2} F. Iocco, Astrophys. J. 677 (2008) L1.
\bibitem{Freese2} K. Freese, D. Spolyar and A. Aguirre, 2008 [arXiv:0802.1724].
\bibitem{Ekstrom Maynet} S. Ekstr$\ddot{\mbox{o}}$m, G. Meynet, A. Maeder
and F. Barblan, Astronomy and Astrophysics 478 (2008) 467.
\bibitem{Gould} A. Gould, ApJ 567 (1987) 532.
\bibitem{SI}
  Z.~Ahmed {\it et al.}  [CDMS Collaboration],
  arXiv:0802.3530 [astro-ph];
  J.~Angle {\it et al.}  [XENON Collaboration],
  Phys.\ Rev.\ Lett.\  {\bf 100} (2008) 021303
  [arXiv:0706.0039 [astro-ph]].
\bibitem{SD}
  J.~Angle {\it et al.},
  arXiv:0805.2939 [astro-ph];
  E.~Behnke {\it et al.}  [COUPP Collaboration],
  Science {\bf 319} (2008) 933
  [arXiv:0804.2886 [astro-ph]].
\bibitem{GriestSeckel} K.Griest and D.Seckel, Nucl. Phys. B 296 (1987)
681.
\bibitem{Freese3} K. Freese, P. Gondolo, J.A. Sellwood and D. Spolyar, [arXiv:0805.3540].
\bibitem{AbelBryan}  T. Abel, G.L. Bryan and M.L. Norman, Science
295 (2002) 93.
\bibitem{Yoon} S. Yoon, F. Iocco and S. Akiyama 2008 [arXiv:0806.2662].

\end{thebibliography}
\end{document}